\documentclass[aps,pra,twocolumn,nofootinbib]{revtex4}
\usepackage{bm}
\usepackage{graphicx,graphics,psfrag}
\usepackage{amsbsy}
\usepackage{amsmath}
\usepackage{amsfonts}
\usepackage{amsthm}
\usepackage{color}

\newcommand{\red}{\color{red}}

\newcommand{\black}{\color{black}}

\begin{document}
\title{Measuring the quality of a quantum reference frame: the relative entropy of frameness}
\author{Gilad Gour}
\email{gour@math.ucalgary.ca}
\affiliation{Institute for Quantum Information Science and Department of
Mathematics and Statistics, University of Calgary, 2500 University Drive NW, Calgary, Alberta, Canada T2N 1N4}
\author{Iman Marvian}
\email{imarvian@uwaterloo.ca} \affiliation{Institute for Quantum Computing, University of Waterloo, 200 University Ave. W, Waterloo, Ontario, Canada N2L 3G1}
\author{Robert W. Spekkens}
\email{rspekkens@perimeterinstitute.ca} \affiliation{Perimeter Institute for Theoretical Physics, 31 Caroline
St. N, Waterloo, Ontario, Canada N2L 2Y5}
\date{Apr. 17, 2009}

\begin{abstract}
In the absence of a reference frame for transformations associated with a group G, any quantum state that is
non-invariant under the action of G may serve as a token of the missing reference frame. We here introduce a
novel measure of the quality of such a token: the relative entropy of frameness. This is defined as the relative
entropy distance between the state of interest and the nearest G-invariant state. Unlike the relative entropy of
entanglement, this quantity is straightforward to calculate and we find it to be precisely equal to the
G-asymmetry, a measure of frameness introduced by Vaccaro \emph{et al.} It is shown to provide an upper bound on
the mutual information between the group element encoded into the token and the group element that may be
extracted from it by measurement. In this sense, it quantifies the extent to which the token successfully
simulates a full reference frame. We also show, that despite a suggestive analogy from entanglement
theory, the regularized relative entropy of frameness is zero and therefore does not quantify the rate of interconversion between the
token and some standard form of quantum reference frame.  Finally, we show how these investigations
yield a novel approach to bounding the relative entropy of entanglement.
\end{abstract}

\pacs{03.67.Mn, 03.67.Hk, 03.65.Ud}
\maketitle

\theoremstyle{plain} \newtheorem{theorem}{Theorem} %
\newtheorem{lemma}[theorem]{Lemma} \newtheorem{corollary}[theorem]{Corollary}
\newtheorem{conjecture}[theorem]{Conjecture} %
\newtheorem{proposition}[theorem]{Proposition}

\theoremstyle{definition} \newtheorem{definition}{Definition}

\theoremstyle{remark} \newtheorem*{remark}{Remark} %
\newtheorem{example}{Example}

\section{Introduction}

Transformations are defined relative to reference frames. For instance, a rotation can only be implemented
relative to a physical system -- such as a gyroscope -- that defines the axis of rotation.  When a
reference frame for some particular group of transformations is lacking, a quantum state that is non-invariant
under the action of the group may serve as a token of it, allowing one to emulate operations that would normally
require the reference frame. This is analogous to the manner in which an entangled state can stand in for a
quantum channel through the teleportation protocol. Indeed, just as an entangled state is a resource, so too is
a quantum token of a reference frame and to make best use of it one must determine how it is interconverted from
one form to another and distilled into a standard form~\cite{SVC04,Enk05,GS07}, how it is exploited to perform
non-invariant operations~\cite{Boi08,Mar08}, and how it degrades with use~\cite{Bar06,Pou06}. In this article,
we seek to quantify the quality of particular quantum states as tokens of a reference frame.

To be operational, a measure of frameness must be monotonically non-increasing under $G$-invariant
operations in which case it is called a $G$\emph{-frameness monotone}~\cite{GS07} (see also appendix A
of Ref.~\cite{Bar06}).  There are many types of such measures. One type quantifies the extent to which tasks
requiring the reference frame can be implemented using only the quantum token. Because it measures the ability
of the quantum sample to simulate the classical reference frame, we call this a \emph{simulation measure}.The
second quantifies the degree to which the quantum sample can be converted to (or obtained from) some standard
form.  Measures of this type will be called \emph{conversion measures.} A preliminary investigation of
these was undertaken in Ref.~\cite{GS07}. Finally, one can consider a more abstract approach, the
operational significance of which is not clear \emph{a priori}: define a \emph{geometric} measure of frameness
by the distance of a given frame state to the nearest $G$-invariant state. We shall have something to say about
each type of measure.

Our focus is the analogue for reference frames of the relative entropy of entanglement \cite{Ved98}. The latter
is a geometric measure of entanglement for mixed states, specifically, the relative entropy distance between an
entangled state and the nearest separable state. By analogy, we define the \emph{relative entropy of
}$\emph{G-}$\emph{frameness }to be the relative entropy distance between a frame state and the nearest $G$
-invariant state. Whereas the problem of finding an explicit formula for the relative entropy of entanglement is
extremely difficult (indeed, it remains an open problem even in the case of two qubits \cite{Eisert}), we find
that the relative entropy of frameness is easy to calculate. In fact, it turns out to be precisely equal to the
$G$-asymmetry of Vaccaro \emph{et al}.~\cite {Vac05}, defined as the difference between the von Neumann entropy
of the $G$-twirled state and that of the state itself.  This is the main result of
Sec.~\ref{sec:GframenessGasymmetry}.  We calculate the $G$-asymmetry explicitly in a few special cases.

Vaccaro \emph{et al.} have shown that the $G$-asymmetry provides a tight upper bound on the amount of work that
can be extracted from the quantum token of the reference frame. We demonstrate in
Sec.~\ref{sec:signif4simulation} that it also has operational significance as a simulation measure, providing an
upper bound on the extent to which the token can encode information about a group element.

We also show that the G-asymmetry is relevant for conversion measures. This might be expected from analogy
with entanglement theory. We know that the regularized relative entropy of entanglement is equal to the entanglement
of distillation for a set of states that can be reversibly transformed into pure states by Local Operations
and Classical Communications (LOCC)~\cite{HOH02}, and in particular for pure bipartite states it is equal to the entropy of entanglement.
Furthermore, the results of Horodecki~\emph{et al.}~\cite{HOH02} suggest that for a class of states that can be
reversibly transformed from one state to another by a set of allowed operations, the regularized relative entropy
distance to a set of non-resource states always quantifies the rate of distillation to a standard form of the
resource.~\footnote{Indeed, the fact that this suggestion contradicts the results of our previous work on
distilling a standard form of reference frame \cite{GS07} motivated some of this work.}

However, in Sec.~\ref{sec_asym} we show that the regularized relative entropy of frameness is always zero.
This happens because the relative entropy of $G$-frameness, the most natural geometric measure of $G$-frameness
and a useful simulation measure of $G$-frameness, is not an extensive quantity in the thermodynamic sense. By
contrast, the relative entropy of entanglement \emph{is} an extensive quantity. This fundamental difference
between the resource theory of quantum reference frames and the resource theory of entanglement is likely to be
significant for making sense of other differences one finds when comparing the two sorts of resources.

Although its regularization is always zero, we conjecture that the relative entropy of frameness still has relevance for conversion measures. We discuss this possibility in Sec.~\ref{sec:speculation}.

The ease with which one can compute the relative entropy of frameness is a consequence of the fact that there is
an operation (the $G$-twirling operation) that maps all states to states that have no frameness (the
$G$-invariant ones). Insofar as this feature might be reproduced in other resource theories, we expect that a
geometric measure of such resources
---the relative entropy distance to non-resource states-- might be similarly easy to calculate. Specifically,
we demonstrate via Theorem~\ref{theorem} that if the set of non-resource states is the image of some operation
$\mathcal{E}$ acting on the full set of states, and furthermore $\mathcal{E}$ is both a unital and idempotent
superoperator, then the relative entropy distance of an arbitrary state $\rho$ to the set of non-resource states
can be easily calculated: it is simply the relative entropy distance between $\rho$ and $\mathcal{E}(\rho)$.
Even if a resource theory fails to have this feature, some insight may be gained into geometric measures of the
resource.  For instance, in the case of entanglement theory, one might expect to find some interesting upper
bounds on the relative entropy of entanglement by identifying operations that map all states to separable ones.
In Sec.~\ref{sec:ree} we show that this is indeed the case by identifying some operations of this sort. The
bounds we obtain in this way are found to be tight in some cases.

\section{The relative entropy of G-frameness and the G-asymmetry} \label{sec:GframenessGasymmetry}

Let $G$ be a finite or compact Lie group with a unitary representation $T:G\rightarrow
\mathcal{B}(\mathcal{H})$ where $\mathcal{B}(\mathcal{H})$ denotes the bounded operators on the Hilbert space
$\mathcal{H}$.  Let $\mathcal{S}(\mathcal{H})$ be the set of normalized states.  The set of $G$-invariant states
is denoted by $\mathrm{INV}(G)$,
\begin{equation*}
\mathrm{INV}(G)\equiv \{\sigma |\forall g\in G:T(g)\sigma T^{\dag }(g)=\sigma , \sigma \in
\mathcal{S}(\mathcal{H})\}.
\end{equation*}
These are clearly the only states that can be prepared by someone who lacks a reference frame for
transformations associated with the group $G$.  For instance, if one does not have a physical system to define
``up along the $\hat{z}$-axis'', then it is impossible to prepare a spin 1/2 system in an eigenstate of angular
momentum along the $\hat{z}$-axis.  The fact that the restriction of lacking a reference frame takes this form
(that of a superselection rule) is discussed at length in previous work, such as Sec.~II of Ref.~\cite{BRS07}
and Sec.~II.A of Ref.~\cite{GS07}, to which we refer the reader.

It is useful to note two other ways in which the set of $G$-invariant states may be characterized. Let
$\mathcal{G}:\mathcal{B}(\mathcal{H})\to \mathcal{B}(\mathcal{H})$ be the trace-preserving completely positive
linear map
\begin{equation}
\mathcal{G}[\rho ]\equiv \int_{G}\text{d}g\,T(g)\rho T^{\dag }(g), \label{eq:Gaveraging}
\end{equation}%
which averages over the action of the group $G$ with the $G$-invariant (Haar) measure \textrm{d}$g$. (For a
finite group, one simply replaces the integral with a sum.) $\mathcal{G}$ is called the $G$\emph{-twirling
operation}.

The set of $G$-invariant states is equivalent to the set of states that are fixed points of
$\mathcal{G}$,
\begin{eqnarray}
\label{eq:Fix} \mathrm{INV}(G) &=&\mathrm{Fix}(\mathcal{G})  \\
 &\equiv&\{\sigma |\mathcal{G}(\sigma
)=\sigma,\sigma \in \mathcal{S}(\mathcal{H})\}. \nonumber
\end{eqnarray}
This is easily verified to be a consequence of the invariance of the measure.

The set of $G$-invariant states is also equivalent to the image of $\mathcal{G}$,
\begin{eqnarray}
\label{eq:Image}
\mathrm{INV}(G) &=&\mathrm{Image}(\mathcal{G})  \\
&\equiv&\{\sigma |\sigma =\mathcal{G}(\rho ),\rho \in \mathcal{S}(\mathcal{H})\}. \nonumber
\end{eqnarray}
To see this, we make use of the following fact, the proof of which is straightforward.
\begin{lemma}
\label{lemma:FixImage}
For a map $\mathcal{E}$,
$\mathrm{Image}(\mathcal{E})=\mathrm{Fix}(\mathcal{E})$ if and only if $\mathcal{E}^2 = \mathcal{E}$ (that is,
$\mathcal{E}$ is idempotent).
\end{lemma}
Given that the $G$-twirling operation $\mathcal{G}$ is idempotent (this follows trivially from the invariance of
the measure), we infer from this lemma that $\mathrm{Image}(\mathcal{G})=\mathrm{Fix}(\mathcal{G})$, and
consequently Eq.~(\ref{eq:Fix}) implies Eq.~(\ref{eq:Image}).  We are now in a position to define our measure of
frameness.

Recall that the relative entropy distance between $\sigma $ and $\rho $ is
\begin{align}
S\left( \rho \|\sigma \right) & \equiv \mathrm{Tr}(\rho \log \rho )-\mathrm{%
Tr}(\rho \log \sigma ) \\
& =-S(\rho )-\mathrm{Tr}(\rho \log \sigma ),
\end{align}%
where $S$ denotes the von Neumann entropy and where all the logarithms  are in base 2.
\begin{definition}
the \emph{relative entropy of }$G$\emph{-frameness} of a state $\rho \in \mathcal{S}(\mathcal{H})$ is the
relative entropy distance of $\rho$ to the nearest $G$-invariant state,
\begin{equation*}
R_{\mathrm{INV}(G)}(\rho )=\min_{\sigma \in \mathrm{INV}(G)}\{S\left( \rho \Vert \sigma \right) \}.
\end{equation*}
\end{definition}

Vaccaro \emph{et al. }\cite{Vac05} introduced the following measure of frameness:
\begin{definition}
The $G$-\emph{asymmetry }of a state $\rho  \in \mathcal{S}(\mathcal{H})$ is
\begin{equation*}
A_{G}(\rho )\equiv S(\mathcal{G}(\rho ))-S(\rho ),
\end{equation*}
where $S$ denotes the von Neumann entropy.
\end{definition}

The G-asymmetry was proven to be a $G$\emph{-frameness monotone }in Ref.~\cite{Vac05} (more precisely, it was
shown to be, in the terminology of Ref.~\cite{GS07}, an \emph{ensemble} monotone). This confirms that it is a
good measure of frameness. It also has the nice feature of being nonnegative and equal to zero for $G$-invariant
states.

Our first result is the following.
\begin{proposition} \label{proposition}
The relative entropy of G-frameness is equal to the G-asymmetry and the
G-invariant state with the smallest relative entropy distance to $%
\rho $ is $\mathcal{G}(\rho )$,
\begin{eqnarray*}
\min_{\sigma \in \mathrm{INV}(G)}S\left( \rho \Vert \sigma \right)  &=&S(%
\mathcal{G}(\rho ))-S(\rho ) \\
&=&S\left( \rho \Vert \mathcal{G}(\rho )\right) \,.
\end{eqnarray*}
\end{proposition}

Special cases of this result have been derived in previous work.  Specifically, \.{A}berg's \emph{relative
entropy of superposition} \cite{Abe07}, which seeks to quantify the degree of superposition relative to an
orthogonal decomposition of the Hilbert space, is equivalent to the relative entropy of U(1)-frameness where the
irreducible representations of U(1) pick out the orthogonal decomposition.  The results of Ref.~\cite{Abe07}
therefore imply proposition \ref{proposition} for the case of G=U(1). In the further specialized case where
there is no multiplicity in the representations of U(1)
the result was proven by Horodecki~\emph{et al.}~\cite{Hor05}.

Proposition \ref{proposition} is itself a special case of a much more general result which we shall present as a
theorem. In Sec.~\ref{sec:ree}, it will be used to provide a bound on the relative entropy of entanglement.

To state and prove the theorem we recall some relevant definitions and properties of a quantum channel
$\mathcal{E}:\mathcal{S}(\mathcal{H})\rightarrow \mathcal{S}(\mathcal{H})$. The adjoint of a superoperator is
defined by the Hilbert-Schmidt inner product, specifically, \textrm{Tr}$(A\mathcal{E}(B))=$
\textrm{Tr}$(\mathcal{E}^{\dag }(A)B)$ for all $A,B\in \mathcal{S}(\mathcal{H}).$ The channel $\mathcal{E}$ is
called unital if $\mathcal{E}(I)=I$.  By the results of Ref.~\cite{Kri03} we know that the fixed point set of a
unital channel is an algebra.  Specifically, if the quantum channel has Kraus decomposition
$\mathcal{E}(\rho)=\sum_a E_a \rho E_a^\dag$, then the fixed point set of $\mathcal{E}$ is the algebra
$\mathcal{A}$ given by the commutant of $\{E_a,E_a^\dag\}$,
\begin{equation}
\mathcal{A}=\{\tau \in \mathcal{B}(\mathcal{H}):\  [\tau, E]=0 \ \  \forall E\in\{E_a,E_a^\dag\}\  \}
\end{equation}
This means that the fixed point set for a unital trace-preserving quantum operation $\mathcal{E}$ is equivalent
to that of $\mathcal{E}^\dag$. Moreover we can deduce that for $\tau \in
\mathrm{Fix}(\mathcal{E})$
\begin{equation}
\mathcal{E}^\dag(\tau^n) = \tau^n.   \label{eq:powers}
\end{equation}
for all integers  $n$.

\begin{theorem} \label{theorem}
Suppose $\mathcal{E}$ is a trace-preserving completely positive map that satisfies the following two properties:
(i) It is unital
 and (ii) It is idempotent, or equivalently (by lemma \ref{lemma:FixImage}), every state in the
image of $\mathcal{E}$ is a fixed point of $\mathcal{E}$,
\begin{equation}
\textrm{Image}(\mathcal{E})=\textrm{Fix}(\mathcal{E}).
 \label{eq:property2}
\end{equation}
In this case, the minimum relative entropy distance between an arbitrary state $\rho \in \mathcal{S}(\mathcal{H})$ and a state
$\sigma \in \mathrm{Image}(\mathcal{E})$ satisfies
\begin{eqnarray}
\min_{\sigma \in \mathrm{Image}(\mathcal{E})}S\left( \rho \Vert \sigma
\right)  &=&S(\mathcal{E}(\rho ))-S(\rho )  \label{eq:theoremline1} \\
&=&S\left( \rho \Vert \mathcal{E}(\rho )\right) \,.  \label{eq:theoremline2}
\end{eqnarray}
\end{theorem}

\begin{proof} By the definition of the relative entropy,
\begin{equation}
\min_{\sigma \in \mathrm{Image}(\mathcal{E})}S\left( \rho \Vert \sigma
\right) =-S(\rho )+\min_{\sigma \in \mathrm{Image}(\mathcal{E})}\left[ -%
\mathrm{Tr}\left( \rho \log \sigma \right) \right] .
\end{equation}%
The proof of Eq.~(\ref{eq:theoremline1}) then proceeds in two steps. \ First, it is shown that for $\rho \in
\mathcal{S}(\mathcal{H})$ and $\sigma \in \mathrm{Image}(\mathcal{E})$,
\begin{equation}
\mathrm{Tr}(\rho \log \sigma )=\mathrm{Tr}(\mathcal{E}(\rho )\log \sigma )\,. \label{eq:step1}
\end{equation}%
Second, it is shown that
\begin{equation}
\min_{\sigma \in \mathrm{Image}(\mathcal{E})}\left[ -\mathrm{Tr}\left( \mathcal{E}(\rho )\log \sigma \right)
\right] =S(\mathcal{E}(\rho )) \label{eq:step2}
\end{equation}

We begin by establishing Eq.~(\ref{eq:step1}). By definition of the adjoint of a superoperator we have  %
\begin{equation*}
\mathrm{Tr}(\mathcal{E}(\rho )\log \sigma ) =\mathrm{Tr}(\rho \mathcal{E}%
^{\dag }\left( \log \sigma \right) ) .
\end{equation*}
It follows from Eq.~(\ref{eq:powers}) that for $\sigma \in \mathrm{Image}(\mathcal{E})$ and for any
analytic function $f$,
\begin{equation*}
\mathcal{E}^\dag\left[ f(\sigma )\right] =f(\sigma ).
\end{equation*}
Recalling that the function $\log (1-x)$ is analytic for $0\leq x<1$, it follows that $\log (I-X)$ is analytic
if the operator $X$ satisfies $0\leq X<I.$ \ Given that $0\leq I-\sigma \leq I$, it follows that
\begin{equation*}
\mathcal{E}^\dag\left[ \log \sigma \right] =\log \sigma ,
\end{equation*}
which concludes the first step of the proof.

To demonstrate Eq.~(\ref{eq:step2}), it suffices to note that by Klein's
inequality \cite{Nie00}, the relative entropy distance is non-negative,%
\begin{equation*}
S(\mathcal{E}(\rho )||\sigma )\geq 0,
\end{equation*}
with equality achieved if and only if $\sigma =\mathcal{E}(\rho )$, so that
\begin{equation*}
\min_{\sigma \in \mathrm{Image}(\mathcal{E})}S(\mathcal{E}(\rho )||\sigma )=0.
\end{equation*}%
By the definition of the relative entropy, it follows that%
\begin{equation*}
-S(\mathcal{E}(\rho ))+\min_{\sigma \in \mathrm{Image}(\mathcal{E})}\left[ -%
\mathrm{Tr}\left( \mathcal{E}(\rho )\log \sigma \right) \right] =0\,,
\end{equation*}%
which establishes Eq.~(\ref{eq:step2}) and concludes the proof of Eq.~(\ref{eq:theoremline1}).

\strut Equation (\ref{eq:theoremline2}) is shown as follows. \ By the definition of the relative entropy,
\begin{equation*}
S\left( \rho \Vert \mathcal{E}(\rho )\right) =-S(\rho )-\mathrm{Tr}\left( \rho \log \mathcal{E}(\rho )\right) .
\end{equation*}%
But applying Eq.~(\ref{eq:step1}) with $\sigma = \mathcal{E}(\rho )$, we have
\begin{equation*}
\mathrm{Tr}\left( \rho \log \mathcal{E}(\rho )\right) =\mathrm{Tr}\left( \mathcal{E}(\rho )\log \mathcal{E}(\rho
)\right)=-S(\mathcal{E}(\rho)) .
\end{equation*}
\end{proof}

Proposition \ref{proposition} is a corollary of theorem \ref{theorem} because the G-twirling operation satisfies
both of the requisite conditions of the theorem. Property (i), that $\mathcal{G}$ is unital, is trivial to
see, \black and property (ii), that $\mathcal{G}^2 = \mathcal{G}$, follows from the invariance of the measure
$dg$ in $\mathcal{G}$, as noted below lemma \ref{lemma:FixImage}.

We compute the relative entropy of frameness in several simple examples in Appendix \ref{App:A}.

\section{Significance for simulation measures of frameness} \label{sec:signif4simulation}

Vaccaro \emph{et al. }demonstrated that the $G$-asymmetry of a state $\rho $ has the following
operational significance: it provides a tight upper bound on the thermodynamic work that can be extracted from
$\rho $ (with the help of another state) \cite{Vac05}. We shall demonstrate that it also provides a bound on the
state's information content about the group $G$.

Consider that the most common use to which one would put a quantum sample of a reference frame is the task of
estimating the relative orientation of a pair of reference frames. Here the quantum state is prepared relative
to one reference frame and is measured relative to another. The task is to gain information about the group
element describing the relative orientation of the two reference frames.

The estimator is faced with distinguishing states $\{\rho (g)|g\in G\}$ where $\rho (g)=T(g)\rho T^{\dag }(g)$.
The measurement is denoted $E:G\rightarrow \mathcal{P}(\mathcal{H})$, where $ E(g^{\prime })\mathrm{d} g^{\prime
} \geq 0$ and $\int E(g^{\prime }) \mathrm{d}g^{\prime }=I$ with \textrm{d}$g$ the $G$-invariant measure.

The figure of merit for the task can be defined in terms of the probability density $p(g^{\prime
}|g)=\mathrm{Tr}(\rho (g)E(g^{\prime }))$ associated with estimating that the relation is $g^{\prime }$ when the
actual relation is $g.$ Typically, the figure of merit has been defined in terms of a cost
function, leading, for instance, to a consideration of the fidelity between $%
g^{\prime }$ and $g.$ However, another natural measure of how much information has been gained about the group
element is the mutual
information between $g^{\prime }$ and $g,$%
\begin{equation*}
H(g^{\prime }:g)=\int \mathrm{d}g \mathrm{d}g^{\prime }p(g^{\prime },g)\log \frac{p(g^{\prime },g)}{p(g^{\prime
})p(g)},
\end{equation*}%
where $p(g^{\prime },g)$ is the joint probability density of preparing $g$ and estimating $g^{\prime }.$The
\emph{accessible information }is the maximum of the mutual information in a variation over the choice of
measurement,%
\begin{equation*}
\max_{E}H(g^{\prime }:g).
\end{equation*}%
This is simply the classical channel capacity for a channel that randomizes over the action of the group $G$,
but where the variables are continuous rather than discrete. Holevo has provided an upper bound on the quantum
channel capacity for the case of discrete variables \cite{Hol98}, which is readily generalized to the case of
continuous variables~\cite{Bra02}. In our case, it yields
\begin{align*}
\max_{E}H(g^{\prime }& :g)\leq S\left( \int \mathrm{d}g T(g)\rho
T^{\dag }(g)\right)  \\
& -\int \mathrm{d}g S(T(g)\rho T^{\dag }(g))
\end{align*}%
But given that the entropy is invariant under unitaries, $S(T(g)\rho T^{\dag
}(g))=S(\rho ),$ and making use of the $G$-twirling operation of Eq.~(\ref%
{eq:Gaveraging}), we find that the Holevo bound is
\black
simply the $G$-asymmetry \footnote{ The equality of the
$G$-asymmetry and the Holevo $\chi $ quantity was noted in Ref. \cite{Vac05}.},
\begin{equation*}
\max_{E}H(g^{\prime }:g)\leq S\left( \mathcal{G}[\rho ]\right) -S(\rho ).
\end{equation*}
Consequently, the $G$-asymmetry of a state $\rho $ provides an upper bound
on the amount of information about the reference frame that can be encoded
in $\rho .$

\section{Significance for conversion measures of frameness } \label{sec:insignif4conversion}

\subsection{Review of conversion measures of frameness}

We begin by reviewing what is known about conversion measures of frameness. In Ref.~\cite{GS07}, several
conversion measures were defined.  Three sorts of manipulations of frame states were considered: single-copy
deterministic transformations, single-copy stochastic transformations, and asymptotic deterministic
transformations (i.e. transformations among many copies in the limit that the number of copies is infinite).
These were considered only for pure states and for the groups Z$_{2},$ U(1) and SU(2). \footnote{ We are only
interested in frame states as unipartite resources. The resource theory arising from the restriction of two
parties having neither local nor shared reference frames and only able to implement local operations and
classical communication has been considered by several authors~\cite{SVC04,Enk05}, and some have proposed
measures of the degree to which quantum states can stand in for a shared reference, but we shall not consider
this case here.}.

We will focus here on asymptotic interconversion of frame states.  We begin with a few general comments on
asymptotic rates of conversion.  In fact, these comments apply to any resource theory, and consequently, we
state them in a generic form. For an arbitrary pair of resource states, one may find that interconversion,
though possible, cannot be achieved reversibly. In other words, one might find that the conversion rate in one
direction is not the inverse of the conversion rate in the opposite direction. In this case, one can distinguish
the rate at which one can produce a ``gold standard'' resource state from the given state (the amount of the
standard that can be distilled from the given state) and the rate at which one can produce the given state from
that standard (the cost of the given state, in terms of the standard). However, one can always classify the
resource states into classes, such that reversible asymptotic interconversion is possible \emph{within} but not
\emph{between} the classes. Within each such class, one can choose a particular state as the standard and the
rate at which one can convert any state in the class into this standard form becomes a unique measure of
frameness (from which any other rate of asymptotic interconversion among states in the class can be determined).
The most simple resource theories are those for which there is only a single class, that is, all state are
reversibly interconvertible one to the other.

As a concrete example, in entanglement theory, the pure bipartite entangled states form a single class: any
entangled state can be converted asymptotically into any other.  However, the pure tripartite entangled states
are divided into many classes.

For quantum reference frames, something similar occurs. As shown in Ref.~\cite{GS07}, as one varies the nature
of the group, one varies the number of classes within which reversible interconversion of pure frame states is
possible. There is only a single class for the group Z$_{2}$, but many classes for U(1) and SU(2).

Furthermore, the unique measure of frameness within some of these classes has been determined in
Ref.~\cite{GS07}. For instance, one class of U(1)-noninvariant states within which reversible asymptotic
interconversion is the class of states with a gapless number spectrum.  Here, the variance over number is the
unique conversion measure of U(1)-frameness within this class, from which the rate of interconversion between
any two can be computed by taking the ratio. This concludes the review material.

\subsection{A suggestive but misleading analogy}

The question we seek to address in this section is whether the relative entropy of frameness is relevant to
conversion measures of frameness. That this might be the case is suggested by analogy to some results from
entanglement theory. In particular, in Ref.~\cite{HOH02} it is shown that
the \emph{regularized} version of the relative entropy of
entanglement is the unique measure of entanglement.
By analogy, one might expect that for classes of states within which reversible interconversion is possible, the
regularized version of the relative entropy of frameness might be the unique measure of frameness, from which
all rates of interconversion can be inferred.

At first glance, this idea seems to be supported by Ref.~\cite{HOH02} because the relevant result therein is
supposed to be true for all resource theories, not just entanglement theory. Specifically, the authors seem to
show that for an arbitrary resource theory, the regularized relative entropy distance from the given resource
state to the set of non-resource states gives the unique measure of a resource within a class of states wherein
reversible interconversion is possible.

This suggests that in the case of the reference frame resource theory the unique measure of frameness is
obtained by
\begin{equation}
A_{G}^{\infty}(\rho^{\otimes N})=\lim_{N\rightarrow\infty}\frac{1}{N}A_G(\rho^{\otimes N})
\end{equation}
However, as we will show below, the above expression is always zero for finite and compact Lie groups and
consequently cannot be used to compute the rate of interconversion among states.

The moral of the story is that the result of Ref.~\cite{HOH02} only applies if the regularized relative entropy is nonzero and finite.  This constraint was not made explicit in Ref.~\cite{HOH02} and this produced the mistaken impression that the regularized relative entropy must always be the unique measure of a resource.

Nonetheless, the relative entropy of a resource may still have some significance for conversion measures. For instance, as we will show in Sec. \ref{sec:speculation}, in the case of a phase reference one obtains the asymptotic rate of interconversion by regularization of a nonlinear function of the relative entropy of U(1)-frameness.

In order to clarify what conclusions can and cannot be drawn from the results of Ref.~\cite{HOH02}, we begin by providing a detailed review of the latter.

\subsection{Rederivation of the result of Ref.~\cite{HOH02}} \label{sec:insignificance4df}

Given a measure of a resource $f(\rho)$, its regularized version is
defined by
\[
f^{\infty}(\rho)=\lim_{N\rightarrow\infty}\frac{1}{N}f(\rho^{\otimes N}).
\]
The regularized version of a resource is always additive; simply note that
\begin{align*}
f^{\infty}(\rho^{\otimes2}) & =\lim_{N\rightarrow\infty}\frac{1}{N}%
f(\rho^{\otimes2N}) =\lim_{M\rightarrow\infty}\frac{2}{M}f(\rho^{\otimes M})\\
&=2f^{\infty}(\rho).
\end{align*}

We will focus now on measures that are asymptotically continuous. Recall
that a function $f$ is asymptotically continuous, if for sequences $\rho_{n},\sigma_{n}$ of states on Hilbert
space $\mathcal{H}_{n}$, $\lim_{n\rightarrow\infty}\left\Vert \rho_{n}-\sigma_{n}\right\Vert _{1}\rightarrow0$
implies~\cite{Ple07}
\[
\lim_{n\rightarrow\infty}\frac{f(\rho_{n})-f(\sigma_{n})}{1+\log\left(
\dim\mathcal{H}_{n}\right)  }\rightarrow0.
\]

A resource theory is defined by a set $C$ of operations (those that can be implemented without the resource).
Consider a set $S$ of states that are reversibly interconvertible asymptotically in the resource theory, that
is, for any pair $\rho,\sigma \in S$, there exists an operation $\mathcal{E}\in C$ such that
\begin{equation}
\lim_{N\rightarrow\infty}\left\Vert \mathcal{E}(\rho^{\otimes N} )-\sigma^{\otimes M(N)}\right\Vert
_{1}\rightarrow0,\label{eq:star}
\end{equation}
where $M(N)$ is an integer depending on $N$. Since the states are \emph{reversibly}
interconvertible it implies that there exists an operation $\mathcal{F}\in C$ such that
\begin{equation}
\lim_{N\rightarrow\infty}\left\Vert \mathcal{F}(\sigma^{\otimes M(N)}
)-\rho^{\otimes N}\right\Vert _{1}\rightarrow0.\label{eq:starstar}
\end{equation}

Suppose $f$ is a deterministic monotone relative to operations in $C,$ that is,
\[
\forall\rho\;,\;\forall\mathcal{E}\in C,\text{ }f(\mathcal{E}(\rho))\leq f(\rho).
\]
In particular, this is true for $\rho^{\otimes N}$,
\begin{equation} \label{eq:ssss}
\forall\rho\;,\;\forall\mathcal{E}\in C,\text{ }f(\mathcal{E}(\rho^{\otimes N}))\leq f(\rho^{\otimes N}).
\end{equation}
Suppose further that $f$ is asymptotically continuous.  For sequences of states
given by $\rho_N = \mathcal{E}(\rho^{\otimes N})$ and $\sigma_N =\sigma^{\otimes M(N)}$, we have
$\lim_{N\rightarrow\infty}\left\Vert \rho_{N}-\sigma_{N}\right\Vert _{1}\rightarrow 0$ by virtue of
Eq.~(\ref{eq:star}), and so by asymptotic continuity, we infer that
\[
\lim_{N\rightarrow\infty}\frac{1}{N}\left[  f(\mathcal{E}(\rho^{\otimes N}))-f(\sigma^{\otimes M(N)})\right]  \rightarrow0.
\]
Together with Eq.~(\ref{eq:ssss}), this implies that
\[
\lim_{N\rightarrow\infty}\frac{1}{N}\left[  f(\rho^{\otimes N}) -f(\sigma^{\otimes M(N)})\right]  \geq0.
\]
A similar line of reasoning yields
\[
\lim_{N\rightarrow\infty}\frac{1}{N}\left[ f(\sigma^{\otimes M(N)})- f(\rho^{\otimes N})\right]  \geq0.
\]
and so we conclude that
\[
\lim_{N\rightarrow\infty}\frac{1}{N}\left[  f(\rho^{\otimes N}) -f(\sigma^{\otimes M(N)})\right] =0.
\]
But then it follows that
\begin{align*}
f^{\infty}(\rho)&=\lim_{n\rightarrow\infty} \frac{M(N)}{N}\frac{1}{M(N)}f(\sigma^{\otimes M(N)})\\
&=f^{\infty}(\sigma)\lim_{N\rightarrow\infty} \frac{M(N)}{N}.
\end{align*}
So, if $f^{\infty}(\sigma)\neq0,$ then
$$
\lim_{N\rightarrow\infty}\frac{M(N)}{N}=\frac{f^{\infty}(\rho)}{f^{\infty
}(\sigma)}.
$$
We have therefore proven the following theorem.
\begin{theorem} \label{thm:HO} Consider a class of states among which asymptotic
reversible interconversion by operations in the class $C$ (that do not require the resource) is possible. For
any resource measure $f$ that is a deterministic monotone relative to $C$ and asymptotically continuous, if its regularized version $f^{\infty}$ is nonzero and finite, then
$f^{\infty}$ is the unique measure of the resource (ratios of which determine the rate of interconversion
between any two states in the class).
\end{theorem}

Theorem \ref{thm:HO} can be inferred from the calculations in  Ref.~\cite{HOH02}. But in that work it is
not made explicit that the regularized relative entropy must be nonzero and finite.  The authors did not
anticipate that there could be resource theories where the regularized relative entropy might be zero. However,
as we demonstrate below, the resource theory of quantum reference frames is such a case. Specifically, we show
that for all finite groups and all compact Lie groups, the regularized relative entropy distance to the set of
$G$-invariant states is zero.

\subsection{The relative entropy of frameness in the asymptotic limit}

We now consider the dependence of the relative entropy of frameness on $N$, the number of systems, in the
asymptotic limit.  We begin with a simple example that we solve completely, that of a phase reference.  We then
generalize it to arbitrary groups.

 \subsubsection{Phase reference}

Consider the resource theory of quantum phase references, which transform according to the U(1) group. We
will show that the relative entropy of U(1)-frameness depends logarithmically on $N$, implying that the
regularized relative entropy of U(1)-frameness is zero.

As demonstrated in Ref.~\cite{GS07}, and mentioned in Sec.~\ref{sec:insignif4conversion}, only for certain
subsets of pure states does there exists reversible asymptotic interconversion using U(1)-invariant operations.
Consequently, only within such subsets is it possible to define a standard state and thereby a measure of
U(1)-frameness in terms of the rate of distillation of this standard state.  We will here focus our attention on
one such subset of pure states, those with a \emph{gapless number spectrum}.  These are states of the form
$|\psi\rangle = \sum_n c_n |n\rangle$ (the $|n\rangle$ are eigenstates of the total number operator) where the
weights $|c_n|^2$ are nonzero for values of $n$ in a single interval of the natural numbers. As
demonstrated in Ref.~\cite{GS07}, the rate of distillation in the U(1) case is equal to the number variance. See
Ref.~\cite{GS07} for details.

In Ref.~\cite{GS07} it is shown that for states $|\psi\rangle$ with a gapless number spectrum,
\[
\left\vert \psi\right\rangle ^{\otimes N}=\sum_{n=0}^{dN}\sqrt{r_{n}%
}\left\vert n\right\rangle
\]
where $r_{n}$ is a Gaussian distribution in the limit $N\rightarrow\infty.$%
\strut\ \ The regularized relative entropy is
\[
\lim_{N\rightarrow\infty}\frac{1}{N}\min_{\sigma\in\mathrm{U(1)-INV}%
}S(\left\vert \psi\right\rangle ^{\otimes N}\|\sigma).
\]
But%
\begin{align*}
\min_{\sigma\in\mathrm{U(1)-INV}}S(\left\vert \psi\right\rangle ^{\otimes
N}\|\sigma)  & =S(\mathcal{G}(\left\vert \psi\right\rangle ^{\otimes
N})-(S(\left\vert \psi\right\rangle ^{\otimes N})\\
& =S(\sum_{n=0}^{dN}r_{n}\left\vert n\right\rangle \left\langle n\right\vert
)=H(\{r_{n}\}),
\end{align*}
where $H$ is the Shannon entropy. \ So it suffices to determine the Shannon
entropy for a Gaussian distribution.
Suppose%
\[
r_{n}=\frac{1}{\sqrt{2\pi}\sigma}e^{-\frac{\left(  n-\mu\right)  ^{2}}%
{2\sigma^{2}}}\,.
\]
Then, a straightforward calculation gives
$$
H(\{r_{n}\}) =-\sum_{n}r_{n}\log r_{n}=\log\left(  \sqrt{2\pi}\sigma\right)  +\frac{1}{2}.
$$

Note that $\sigma^2\equiv V(\left\vert
\psi\right\rangle ^{\otimes N})$ where $V$ is the number variance defined by
$$ V(\rho)=\mathrm{Tr}[\rho\hat{N}^2]-\left(\mathrm{Tr}[\rho\hat{N}]\right)^2. $$
The variance is additive, so that $V(\left\vert
\psi\right\rangle ^{\otimes N})=NV(\left\vert \psi\right\rangle ).$
Consequently,
\[
H(\{r_{n}\})=\frac{1}{2}\log\Big(2\pi NV(|\psi\rangle)\Big) +\frac{1}{2}.
\]

We have therefore shown the following.
\begin{lemma} \label{lemma:U1assymetry} The relative entropy of U(1)-frameness for $N$ copies of a U(1)-noninvariant state $|\psi\rangle$ is
\begin{equation}
A_{U(1)}(|\psi\rangle^{\otimes N})=\frac{1}{2}\log\Big(2\pi NV(|\psi\rangle)\Big) +\frac{1}{2},
\end{equation}
which is to say logarithmic in $N.$
\end{lemma}

The obvious corollary is:
\begin{corollary}
The regularized relative entropy of U(1)-frameness is zero,
\begin{equation}
A_{U(1)}^{\infty}(|\psi\rangle^{\otimes N})=\lim_{N\rightarrow\infty}\frac{1}{N}A_{U(1)}(|\psi\rangle^{\otimes
N})=0.
\end{equation}
\end{corollary}

There is another way to see that the regularized relative entropy of U(1)-frameness cannot be equal to the
variance (and consequently cannot quantify the rate of distillation): the former is asymptotically continuous,
while the latter is not.  To see that the regularized relative entropy of U(1)-frameness is asymptotically
continuous, it suffices to note that every relative entropy distance is asymptotically continuous (as shown in
Ref.~\cite{DH99}) and that regularization preserves this property.  It therefore suffices to prove the following
proposition.

\begin{proposition}
The variance is not asymptotically continuous.
\end{proposition}
\begin{proof}
Consider the following two sequences of states on a Hilbert space $\mathcal{H}_n$ of dimension $n>4$:
\begin{align*}
|\psi_n\rangle &=\frac{1}{\sqrt{2}}\left(|0\rangle+|n\rangle\right)\\
|\phi_n\rangle &=\sqrt{\frac{1}{2}-\frac{1}{\sqrt{n}}}|0\rangle+\sqrt{\frac{1}{2}+\frac{1}{\sqrt{n}}}|n\rangle\;.
\end{align*}
Clearly, $\lim_{n\rightarrow\infty}\||\psi_n\rangle \langle \psi_n| -|\phi_n\rangle \langle \phi_n|\|_1=0$.
Now, the variances of these states are given by
$$
V(|\psi_n\rangle)=n^2\;\;\;\text{and}\;\;\;V(|\phi_n\rangle)=n^2-4n\;.
$$
Thus,
\begin{equation}
\lim_{n\rightarrow\infty}\frac{V(|\psi_n\rangle)-V(|\phi_n\rangle)}{\log(n)}=\lim_{n\rightarrow\infty}\frac{n}{\log(n)} \not\rightarrow 0\;\;.\nonumber
\end{equation}
This completes the proof.
\end{proof}

\subsubsection{Reference frame for a general group} \label{sec_asym}
We now consider the asymptotic behavior of the relative entropy of frameness for both finite groups and
compact Lie groups.  We will again demonstrate that this quantity is sublinear in $N$ and therefore regularizes
to zero.

First we consider $G$ to be an arbitrary finite group. \black Denote by $|G|$ the cardinality of the
group and by $g_i$ the elements of the group. The $G$-twirling operation on $\rho^{\otimes N}$ can be written as
$$
\mathcal{G}(\rho^{\otimes N})=\frac{1}{|G|}\sum_{i=1}^{|G|}\left[T(g_i)\right]^{\otimes N}\rho^{\otimes N}
\left[T^{\dag}(g_i)\right]^{\otimes N}
$$
Now, for any given ensemble of states $\{p_i\;,\;\rho_i\}$, the von-Neumann entropy satisfies the following
inequality:
$$
S\left(\sum_{i}p_i\rho_i\right)\leq\sum_{i}p_iS(\rho_i)+H(\{p_i\})\;\;,
$$
where $H(\{p_i\})$ is the Shannon entropy. Hence, we find that
\begin{equation} \label{eq:upbound}
S\left(\mathcal{G}(\rho^{\otimes N})\right)\leq S\left(\rho^{\otimes N}\right)+\log |G|\;\;.
\end{equation}
This establishes our first result.
\begin{lemma}
The relative entropy of G-frameness for a finite group G satisfies
$$
A_G(\rho^{\otimes N})\leq\log |G|\;.
$$
\end{lemma}
Note that the upper bound is independent of $N$ so the regularization of $A_G$ for a finite group is clearly zero.

We now proceed to discuss the case of a compact Lie group. To find an upper bound, we will need to use the
following fact from design theory:
\begin{lemma}\cite{Aidan,Pie05}\label{lem1}
Given a group $G$ with a unitary representation $U$ of dimension $d^*$, there exists a finite set
$\{g_i\}_{i=1}^{m(d^*)}$, and weighting probabilities $\{p_i\}_{i=1}^{m(d^*)}$ such that
$$
\int_{G} dg U(g)\sigma U(g^{\dag})=\sum_{i=1}^{m(d^*)}{p_i} U(g_{i})\sigma U(g_{i}^{\dag})
$$
for all states $\sigma$.
\end{lemma}
An upper bound for $m(d^*)$ can be found in Proposition 2.6 of~\cite{Pie05}. Using this result, we conclude
that~\cite{Aidan}:
\begin{equation} \label{up_bound_design}
m(d^*)\leq d^{*2}.
\end{equation}

We are now ready to find an upper bound for $A_G(\rho^{\otimes N})$. Consider the effect of $G$-twirling on the state $\rho^{\otimes N}$.
$$
\mathcal{G}[\rho^{\otimes N}]=\int_{G} dg T(g)^{\otimes N}\rho^{\otimes N} T^{\dag}(g)^{\otimes N}
$$
Assuming $\rho$ is a state in the $d$ dimensional Hilbert space $\mathcal{H}_d$  then  $\rho^{\otimes N}$ lives in the symmetric subspace of $\mathcal{H}_d^{\otimes N}$. By a simple counting argument we find that the dimension of the symmetric subspace is
\begin{equation} \label{eq:dstar}
d^*= {{N+d-1}\choose{d-1}}
\end{equation}
On the other hand, assuming $U(g)=T(g)^{\otimes N}$ then $U$ will be a representation of $G$ which leaves the
symmetric subspace of $\mathcal{H}_d^{\otimes N}$ invariant.  Assuming $\sigma=\rho^{\otimes N}$ and
$U(g)=T(g)^{\otimes N}$ we can use the result of lemma \ref{lem1} to infer that
$$
\mathcal{G}[\rho^{\otimes N}]= \sum_{i=1}^{m(d^*)}{p_i} U(g_{i}) \rho^{\otimes N}  U(g_{i}^{\dag})
$$
and, via Eq.~(\ref{up_bound_design}) and Eq.~(\ref{eq:dstar}),
\begin{equation} \label{upper_bound_m}
m(d^*)\leq {{N+d-1}\choose{d-1}}^2.
\end{equation}

Thus, from Eq.~(\ref{eq:upbound}) we have
$$
S\left(\mathcal{G}[\rho^{\otimes N}]\right)
\leq S(\rho^{\otimes N})+\log
m(d^*)\;.
$$
The G-asymmetry of $\rho^{\otimes N}$ is therefore bounded above by
$$
A_G(\rho^{\otimes N})\leq\log m(d^*)\;.
$$
Now using the upper bound on $m(d^*)$, Eq.~(\ref{upper_bound_m}), and Stirling's approximation we get
\begin{equation}
 \lim_{N\rightarrow\infty} A_G(\rho^{\otimes N}) \leq \lim_{N\rightarrow\infty}2\  \log d^*= 2(d-1) \log N.
 \end{equation}

We summarize this result in a lemma.
\begin{lemma}
In the asymptotic limit $N\rightarrow\infty$, the relative entropy of G-frameness for a compact Lie group G is
bounded above by an expression logarithmic in $N$,
\begin{equation}
A_G(\rho^{\otimes N}) \leq 2(d-1) \log N. \label{asym}
\end{equation}
\end{lemma}

Recall that in Sec.~\ref{sec:signif4simulation}, we found an information theoretic interpretation for the
relative entropy of G-frameness: $A_G(\sigma)$ provides an upper bound on the amount of information that can be
encoded in $\sigma$ about the group element describing the quantum reference frame.  So Eq.(\ref{asym}) means
that the information encoded in $\rho^{\otimes N}$ increases at most logarithmically in $N$.

Note that from the results of this section we can deduce
\begin{corollary} The regularized relative entropy of G-frameness for G a finite or compact Lie group is zero,
\begin{equation}
A_{G}^{\infty}(\rho^{\otimes N})=\lim_{N\rightarrow\infty}\frac{1}{N}A_G(\rho^{\otimes N})=0.
\end{equation}
\end{corollary}

\subsection{Determining the unique measure of a resource when the regularized relative entropy is zero} \label{sec:speculation}
The results thus far give no reason to think that the relative entropy of frameness has any significance at all for conversion measures.  But there is, in fact, a way of obtaining conclusions concerning the latter from the former.

For Lie groups, the fact that the relative entropy of G-frameness is not extensive, i.e. the fact that $A_G(\rho^{\otimes N})$ is not linear in $N$ to leading order, is what blocks us from drawing any interesting conclusions from Theorem \ref{thm:HO}.  An obvious idea, then, is to find a continuous monotonic function $\mathcal{L}:R\rightarrow R$ such that $\mathcal{L}(A_G(\rho^{\otimes N}))$ is linear in $N$ to leading order, so that the monotone $\mathcal{L}(A_G(\cdot))$ does regularize to something finite.

In the case of a phase reference, the relative entropy of U(1)-frameness (equivalently, the U(1)-asymmetry) is given by lemma \ref{lemma:U1assymetry}.  It is logarithmic in $N$.  We can define an extensive monotone by taking $\mathcal{L}(x)=2^{2x}$. We obtain
$$\mathcal{L}(A_{U(1)}(\left\vert \psi\right\rangle^{\otimes N})\ )=4\pi N V(\left\vert \psi\right\rangle)$$
and therefore the regularization of this new monotone yields
\begin{equation}
 \lim_{N\rightarrow\infty}\frac{1}{N} \mathcal{L}(A_{U(1)}(\left\vert \psi\right\rangle^{\otimes N})\ )=4\pi V(\left\vert \psi\right\rangle).
\end{equation}
This is proportional to the variance of $\left\vert \psi\right\rangle$, which is precisely the measure of U(1)-frameness that determines asymptotic rates of conversion, as shown in \cite{GS07}.

This example suggests that for Lie groups one may always be able to infer the unique asymptotic measure of frameness from the relative entropy of frameness.  Indeed, we conjecture that this is the case.  However, to prove that this is the case, one requires a result that is more general than Theorem \ref{thm:HO}, a result that takes into account the Lipschitz constant of the monotone.  We hope to settle this question in future work.

\section{Insights for the relative entropy of entanglement} \label{sec:ree}

An obvious question that arises from our evaluation of the relative entropy of frameness is whether similar
techniques might provide a means of calculating the relative entropy of entanglement. The latter is defined as the relative entropy distance to the set of separable states (which we denote by $\mathrm{SEP}$ ),
\begin{equation*}
R_{\mathrm{SEP}}(\rho )\equiv \min_{\sigma \in \mathrm{SEP}}\{S\left( \rho \Vert \sigma \right) \}.
\end{equation*}
It is an open problem to determine a formula for the relative entropy of entanglement, even in the simplest case of two qubits. We do not solve the problem here,
but merely show how one can obtain interesting new upper bounds on this quantity

The idea is the following. If there is an operation $\mathcal{E}$ that takes the set of all states to a subset
of the separable states, then the relative entropy distance to the nearest state in this subset is an upper
bound on the relative entropy distance to the nearest separable state. Meanwhile, if $\mathcal{E}$ satisfies the
conditions of theorem~\ref{theorem}, then there is a simple formula for the relative entropy distance to this
subset, namely, $S(\mathcal{E}(\rho ))-S(\rho ).$

\begin{theorem} \label{newtheorem}
Suppose $\rho \in \mathcal{S}(\mathcal{H}_A \otimes \mathcal{H}_B)$ is an arbitrary bipartite density operator, $U$ is an arbitrary unitary operator acting on $\mathcal{H}_B$, and $\mathcal{D}_U$ is the dephasing channel along the basis $\{ U|k\rangle \}$ of $\mathcal{H}_B$ (that is, $\mathcal{D}_U(\cdot)\equiv \sum_k U|k\rangle \langle k|U^{\dag}(\cdot)U|k\rangle \langle k|U^{\dag}$). Then the relative entropy of entanglement of $\rho$ satisfies
$$R_{\mathrm{SEP}}(\rho) \leq \min_U  S\left(\  \left[\mathcal{I}\otimes \mathcal{D}_U \right](\rho)\ \right)-S(\rho) $$
\end{theorem}

\begin{proof}
 Define $\mathcal{E}_U(\sigma)=\left[\mathcal{I}\otimes \mathcal{D}_U \right](\sigma)$,
 This channel has the following properties:

\begin{enumerate}
  \item ${\mathcal{E}}_U$ is unital.
  \item $\left(\mathcal{E}_U\right)^2 =\left[\mathcal{I}\otimes \mathcal{D}_U \right]\left[\mathcal{I}\otimes \mathcal{D}_U\right]=\left[\mathcal{I}\otimes \mathcal{D}_U\circ\mathcal{D}_U \right]=\mathcal{E}_U$\\
      where we have used  $\mathcal{D}_U\circ\mathcal{D}_U=\mathcal{D}_U$.
  \item For an arbitrary state $\sigma\in \mathcal{S}(\mathcal{H}_A \otimes \mathcal{H}_B)$ the state $\mathcal{E}_U(\sigma)$ is separable, or in other words $\mathcal{D}_U$ is an entanglement breaking channel.
\end{enumerate}

Because of items 1 and 2, Theorem~\ref{theorem} applies and we can deduce that
$$\min_{\sigma \in \mathrm{Image}(\mathcal{E}_U)}S\left( \rho \Vert \sigma
\right) =S(\mathcal{E}_U(\rho ))-S(\rho ) $$
On the other hand, since the image of $\mathcal{E}_U$ is a subset of separable states, then
$$R_{\mathrm{SEP}}(\rho)=\min_{\sigma \in \mathrm{SEP}}S\left( \rho \Vert \sigma
\right) \leq \min_{\sigma \in \mathrm{Image}(\mathcal{E}_U)}S\left( \rho \Vert \sigma
\right)   $$
So for arbitrary $U$ we have
$$R_{\mathrm{SEP}}(\rho) \leq  S(\mathcal{E}_U(\rho ))-S(\rho )$$
and therefore
$$R_{\mathrm{SEP}}(\rho) \leq \min_U  S\left(\  \left[\mathcal{I}\otimes \mathcal{D}_U\right](\rho)\ \right)-S(\rho) $$
\end{proof}
In the following we apply this upper bound to the case of two qubits; i.e. $\dim{\mathcal{H}_A}=\dim{\mathcal{H}_B}=2$

\subsection{The two qubit case}

In the two qubit case, the dephasing channel $\mathcal{D}_U$ 
can be parameterized by two variables. In particular, we can write the two-qubit version of $\mathcal{E}_U$  in the following form:
\begin{equation*}
\mathcal{E}_{\theta,\gamma }(\rho )=\frac{1}{2}\rho +\frac{1}{2}\left( I\otimes
U_{\theta,\gamma }\right) \rho \left( I\otimes U_{\theta,\gamma }^{\dag }\right)
\end{equation*}%
where
\begin{equation*}
U_{\theta,\gamma }=\left( \cos \theta \right)
\begin{pmatrix}
1 & 0\\
0 & -1
\end{pmatrix}
+ \left( \sin \theta \right)
\begin{pmatrix}
0 & e^{i\gamma}\\
e^{-i\gamma} & 0
\end{pmatrix}\end{equation*}%
Therefore, for the two qubit case our upper bound on the relative entropy of entanglement becomes
$$R_{\mathrm{SEP}}(\rho) \leq \min_{\theta,\gamma}  S\left( \mathcal{E}_{\theta,\gamma }(\rho )\right)-S(\rho). $$
In the following simple example we show that this upper bound can be tight.
\begin{example}
Consider the bipartite mixed state
$$
\rho^{AB}=p|\phi_{+}\rangle\langle\phi_{+}|+(1-p)|\phi_{-}\rangle\langle\phi_{-}|
$$
where $|\phi_{\pm}\rangle\equiv(|00\rangle\pm|11\rangle)/\sqrt{2}$. From the Hashing inequality \cite{Dev05} we know that the
relative entropy of entanglement satisfies
$$
R_{\textrm{SEP}}\left(\rho^{AB}\right)\geq S\left(\rho^A\right)-S\left(\rho^{AB}\right)=1-H_2(\{p\}),
$$
where $H_2(\{p\})=-p\log p-(1-p)\log(1-p)$ is the binary Shannon entropy function. On the other hand, a
straightforward calculation shows that the minimum of the function $S(\mathcal{E}_{\theta,\gamma}(\rho^{AB} ))$ is
obtained at $\theta=\pi/2$ and $\gamma=0$. For these values of $\theta$ and $\gamma$ we have
$$
\mathcal{E}_{\theta=\pi/2,\;\gamma=0}(\rho^{AB} )=\frac{1}{4}
\begin{pmatrix}
2 & 0 & 0 & 0\\
0 & 0 & 0 & 0\\
0 & 0 & 0 & 0\\
0 & 0 & 0 & 2
\end{pmatrix},
$$
and therefore $S(\mathcal{E}_{\theta=\pi/2,\;\gamma=0}(\rho^{AB} ))=1$ which implies
$$
R_{\textrm{SEP}}\left(\rho^{AB}\right)\leq S(\mathcal{E}_{\theta=\pi/2,\;\gamma=0}(\rho^{AB}
))-S(\rho^{AB})=1-H_2(\{p\}).
$$
Combining the Hashing bound with our upper bound gives  $R_{\textrm{SEP}}\left(\rho^{AB}\right)=1-H_2(\{p\})$.
\end{example}

\section{Outlook}

A natural geometric measure of the quality of a resource state is the distance between it and the nearest
nonresource state. Operations that map resources to nonresources seem to provide a useful way of evaluating such
measures. Specifically, we have shown that the relative entropy distance to the nearest G-invariant state -- a
geometric measure of G-frameness -- is expressed simply in terms of the G-twirling operation. Similarly, we
have identified operations for which the image is a subset of the separable states (i.e. operations which are entanglement-breaking on a subsystem) and we have shown how these help to
bound the relative entropy of entanglement. This approach to quantifying a resource appears
to be cognate with attempts to quantify correlations by the amount of noise that one requires to eliminate
them~\cite{Gro05}. It is a topic that warrants further investigation.

There are strong connections between certain geometric and conversion measures of \emph{entanglement}. In
particular the regularized relative entropy of entanglement is equal to the distillable entanglement among
states that are reversibly interconvertible asymptotically.  Nonetheless, the same does not hold true for the
resource theory of quantum reference frames. The regularized relative entropy of frameness is always zero. It
follows that these connections are not generic features of resource theories and intuitions derived from them
may well be misleading. Nonetheless, we may still be able to use the relative entropy of frameness to find out about conversion measures using a different technique. This is another topic for future research.

In addition to the lessons for entanglement theory and resource theories in general, we have drawn several
conclusions concerning the problem of quantifying the quality of a reference frame, and in particular of finding
a measure of the extent to which a token system can simulate a classical reference frame. We have shown that the
G-asymmetry has a very natural operational interpretation in terms of the accessible information in a reference
frame alignment scheme. It therefore provides an alternative to more common figures of merit for alignment
schemes, such as the fidelity between the estimated and the actual orientation (see section V.D.1 of
Ref.~\cite{BRS07}). Furthermore, it nicely captures the intuitive notion that the optimal state to use in such
alignment schemes is the one that has the largest orbit under the action of the group (see section V.A in
Ref.~\cite{BRS07}). We have here considered only reference frames that are local rather than distributed, but
similar techniques should be useful in solving the multipartite version of the problem.

\begin{acknowledgements}
G.G. is grateful to Aidan Roy for discussions and explanations of the bounds given in lemma~\ref{lem1}
and for directing us to Refs.~\cite{Pie05,God89}. R.W.S. is thankful to Jonathan Oppenheim for
many discussions (including one that motivated part of this project in its early stages), to Joan Vaccaro for bringing Ref.~\cite{Abe07} to our attention and to Sarah Croke for pointing us to Ref.~\cite{Bra02}. G.G.
acknowledges support from NSERC and MITACS.  R.W.S. undertook part of this research while at the Department of Applied Mathematics and Theoretical Physics, University of Cambridge, with support from the Royal Society and from the European Union through the Integrated Project QAP (IST-3-015848), SCALA (CT-015714), SECOQC and the QIP IRC(GR/S821176/01), and part of this research while at Perimeter Institute which is supported by the Government of Canada through Industry Canada and by the Province of Ontario through the Ministry of Research and Innovation. I.M. acknowledges support from the Mike and Ophelia Lazaridis Fellowship.
\end{acknowledgements}

\begin{appendix}
\section{States of maximal G-asymmetry} \label{App:A}

The following examples make use of techniques introduced in Ref.~\cite{BRS07}, to which the reader is referred
for more details. Under the action of the unitary representation of a
compact group $G$, a finite dimensional Hilbert space factorizes as follows%
\begin{equation*}
\mathcal{H}=\sum_{q}\mathcal{H}_{q}=\sum_{q}\mathcal{M}_{q}\otimes \mathcal{N%
}_{q}
\end{equation*}%
where $q$ labels the irreps of $G$, $\mathcal{M}_{q}$ is the $q$th representations space, and $\mathcal{N}_{q}$
is the $q$th muliplicity space.
The $G$-twirling operation has the form%
\begin{equation*}
\mathcal{G}\left( \rho \right) =\sum_{q}\mathcal{D}_{\mathcal{M}%
_{q}}\otimes \mathrm{id}_{\mathcal{N}_{q}}\left( \Pi _{\mathcal{H}_{q}}\rho \Pi _{\mathcal{H}_{q}}\right) ,
\end{equation*}%
where $\Pi _{\mathcal{H}_{q}}$ is the projector onto $\mathcal{H}_{q},$ $%
\mathrm{id}_{\mathcal{N}_{q}}$ is the identity map on $\mathcal{N}_{q},$ and $%
\mathcal{D}_{\mathcal{M}_{q}}$ is the completely decohering map on $\mathcal{%
M}_{q}.$ Therefore, for a pure state $\left\vert \psi \right\rangle
\left\langle \psi \right\vert ,$%
\begin{equation*}
\mathcal{G}(|\psi \rangle \langle \psi| )=\sum_{q}p_{q}\frac{\Pi _{\mathcal{M}_{q}}}{%
\dim \mathcal{M}_{q}}\otimes \rho _{\mathcal{N}_{q}}
\end{equation*}%
where $\rho _{\mathcal{N}_{q}}\equiv \frac{1}{p_{q}}\mathrm{Tr}_{\mathcal{M}%
_{q}}\left( \Pi _{\mathcal{H}_{q}}\left\vert \psi \right\rangle \left\langle \psi \right\vert \Pi
_{\mathcal{H}_{q}}\right) $ is the reduced density operator on the multiplicity space $\mathcal{N}_{q}$ and
$p_{q}\equiv \mathrm{Tr}\left( \Pi _{\mathcal{H}_{q}}\left\vert \psi \right\rangle \left\langle \psi \right\vert
\Pi _{\mathcal{H}_{q}}\right) $ is the probability of $q.$

The dimension of the Hilbert space support of $\mathcal{G}(\psi
)$ is bounded above by%
\begin{equation*}
d_{\ast }\equiv \sum_{q}\dim \mathcal{M}_{q}\times d_{q},
\end{equation*}%
where%
\begin{equation*}
d_{q}\equiv \min \left\{ \dim \mathcal{M}_{q},\dim \mathcal{N}_{q}\right\} .
\end{equation*}%
Among states with this support, the maximum entropy is $\log d_{\ast }$ (achieved by the uniform mixture over
the support).Thus, if one can find a pure state $\psi $ such that $S\left( \mathcal{G}(\psi)\right) =\log
d_{\ast },$ then this state achieves the maximum possible $G$-asymmetry. Such a state can indeed be found. It is
\begin{equation} \label{Chiribellastate}
\left\vert \psi \right\rangle =\frac{1}{\sqrt{d_{\ast }}}\sum_{q}\sqrt{\dim \mathcal{M}_{q}\times
d_{q}}\left\vert \psi _{q}\right\rangle
\end{equation}%
where%
\begin{equation*}
\left\vert \psi _{q}\right\rangle =\sum_{k=1}^{d_{q}}\frac{1}{\sqrt{d_{q}}}%
\left\vert \phi _{k}^{(q)}\right\rangle \otimes \left\vert r_{k}^{(q)}\right\rangle ,
\end{equation*}
and where $\left\{ \left\vert \phi _{k}^{(q)}\right\rangle \right\} $ is a
basis for $\mathcal{M}_{q}$ (or a subspace thereof if $d_{q}<\dim \mathcal{M}%
_{q}$), and $\left\{ \left\vert r_{k}^{(q)}\right\rangle \right\} $ is a
basis for $\mathcal{N}_{q}$ (or a subspace thereof if $d_{q}<\dim \mathcal{N}%
_{q}$).  We consider the state of maximal G-asymmetry for two groups of particular interest: U(1) and SU(2).

\textbf{Quantum Phase reference. }Consider an optical phase reference, described by the group U(1).
The unitary representation of U(1) is $U(\phi )=e^{i\phi \hat{N}}$ where $%
\hat{N}$ is the total number operator. The irreps of U(1) are all 1-dimensional and labeled by a nonnegative
integer $n$, corresponding to the eigenvalue of $\hat{N}.$ For an
arbitrary pure state
$$
\left\vert \psi \right\rangle =\sum_{n}\sqrt{p_{n}} \left\vert n\right\rangle ,$$
 where $\left\vert n\right\rangle$ is an
eigenstate of $\hat{N},$ we find the U(1)-asymmetry to be
$$
A_{\mathrm{U(1)} }(\rho )=H(\{p_{n}\}),
$$ where $H(\{p_{n}\})$ is the Shannon entropy of the distribution $p_{n}.$ If the maximum value of $n$ is
$n_{\max },$ the state having $p_{n}=(n_{\max }+1)^{-1}$ achieves the maximum value of the U(1)-asymmetry,
namely, $\log \left( n_{\max }+1\right).$ This state is simply Eq.~(\ref{Chiribellastate}) for the U(1) case.
For an arbitrary
mixed state, the U(1)-asymmetry is found to be $A_{\mathrm{U(1)}}(\rho )=H(%
\mathbf{\{}p_{n}\mathbf{\}})-S(\rho )$ where $p_{n}\equiv \mathrm{Tr}\left( \Pi _{n}\rho \right) $ with $\Pi
_{n}$ the projector onto the $n$th eigenspace of $\hat{N}.$

\textbf{Quantum Cartesian frame. }Consider the case of $N$ spin-1/2 particles under rotations ($N$ is assumed to
be even for simplicity). The unitary
representation of SU(2) of interest is the collective representation $U(%
\mathbf{\theta })=e^{i\mathbf{\hat{J}\cdot \theta }},$ where $\mathbf{\hat{J}%
=}\left( \hat{J}_{x},\hat{J}_{y},\hat{J}_{z}\right) $ is the total angular momentum operator. Under this action,
the Hilbert space factorizes as
follows $\mathcal{H}=\sum_{j=0}^{j_{\max }}\mathcal{H}_{j}=\sum_{j=0}^{j_{%
\max }}\mathcal{M}_{j}\otimes \mathcal{N}_{j}$ where $j_{\max }=N/2,$ the $%
\mathcal{M}_{j}$ are the $(2j+1)$-dimensional representations spaces, and the $\mathcal{N}_{j}$ are the
multiplicity spaces of dimension
$$
\dim\mathcal{N}_{j}={N\choose N/2-j}\frac{2j+1}{N/2+j+1}\;.
$$
Note that $\dim \mathcal{M}_{j}\leq \dim \mathcal{N}_{j}$ for all $%
j<j_{\max }.$ In the exceptional case of the highest irrep $j=j_{\max }$, the multiplicity space is trivial,
that is, $\dim \mathcal{N}_{j_{\max }}=1,$ so we must treat the subspace $\mathcal{H}_{j_{\max }}$ differently
from the others. An arbitrary pure state can be written as
\begin{equation} \label{arbitrarypurestate}
\left\vert \psi \right\rangle =\sum_{j=0}^{j_{\max }}\sqrt{p_{j}}\left\vert \psi _{j}\right\rangle ,
\end{equation}
where $\left\vert \psi _{j}\right\rangle \in \mathcal{H}_{j}$ and for $%
j<j_{\max }$ has the following Schmidt decomposition relative to the
factorization $\mathcal{H}_{j}=\mathcal{M}_{j}\otimes \mathcal{N}_{j},$%
\begin{equation*}
\left\vert \psi _{j}\right\rangle =\sum_{k}\sqrt{q_{k}^{(j)}}\left\vert \phi _{k}^{(j)}\right\rangle \otimes
\left\vert r_{k}^{(j)}\right\rangle ,
\end{equation*}
where the range of $k$ is of cardinality $2j+1.$ SU(2)-twirling yields
\begin{eqnarray*}
\mathcal{G}(\psi ) &=&p_{j_{\max }}\frac{\Pi _{\mathcal{H}%
_{j_{\max }}}}{2j_{\max }+1} \\
&&+\sum_{j=0}^{j_{\max }-1}p_{j}\left( \frac{\Pi _{\mathcal{M}_{j}}}{2j+1}%
\otimes \sum_{k}q_{k}^{(j)}\left\vert r_{k}^{(j)}\right\rangle \left\langle r_{k}^{(j)}\right\vert \right) .
\end{eqnarray*}%
It follows that the SU(2)-asymmetry for an arbitrary pure state is
\begin{eqnarray}
A_{\mathrm{SU(2)}}(\psi ) &=&p_{j_{\max }}\log \left( 2j_{\max }+1\right) \nonumber \\  &&+\sum_{j=0}^{j_{\max
}-1}p_{j}\left[ \log \left( 2j+1\right) +H(\{q_{k}^{(j)}\})\right] \nonumber \\  &&+H(\{p_{j}\}).
\label{starstar}
\end{eqnarray}

The state with the maximum SU(2)-asymmetry is given by Eq.~(\ref{Chiribellastate}) adapted to the group SU(2).
It is the state of the form of Eq.~(\ref{arbitrarypurestate}) that takes
 the distributions $\{q_{k}^{(j)}\}$ to be uniform and $p_{j}\propto \dim
\mathcal{M}_{q}\times \min \left\{ \dim \mathcal{M}_{q},\dim \mathcal{N}%
_{q}\right\} .$ Hence, $p_{j}\propto 2j+1$ for $j=j_{\max }$ and $%
p_{j}\propto (2j+1)^{2} $ otherwise. The maximum SU(2)-asymmetry is simply $A_{%
\mathrm{SU(2)}}=\log \left[ \left( 2j_{\max }+1\right) +\sum_{j=0}^{j_{\max
}-1}(2j+1)^{2}\right] =\log \left[ \frac{4}{3}j_{\max }^{3}+\frac{5}{3}%
j_{\max }+1\right] . $   (Note that one could also obtain these results by optimizing Eq.~(\ref{starstar}) over
$\{p_j\}$ and $\{q^{(j)}_k\}$.)

In the particularly simple case of two spin-1/2 systems, the optimal state
is of the form$\frac{\sqrt{3}}{2}\left\vert \chi _{j=1}\right\rangle +\frac{1%
}{2}\left\vert \psi ^{-}\right\rangle $ where $\left\vert \chi _{j=1}\right\rangle $ is any triplet state, and
the asymmetry is $2.$ \end{appendix}

\end{document}